\documentclass[11pt]{article}
\usepackage{moriond,epsfig}

\def\be{\begin{equation}}
\def\ee{\end{equation}}
\def\bea{\begin{eqnarray}}
\def\eea{\end{eqnarray}}

\begin{document}
\vspace*{4cm}

\title{Standard Model Physics with ATLAS and CMS}

\author{F. PETRUCCI\\
\it{on behalf of the ATLAS and CMS collaborations}
 }

\address{Universit\'a Roma Tre and INFN\\ Via della Vasca Navale, 84 - 00184 Rome, Italy}

\maketitle\abstracts{
The study of Standard Model (SM) physics is crucial at the LHC for several reasons.
Before any discovery can be claimed a detailed understanding
of the detectors should be reached and benchmark SM processes should be 
measured. A precise measurement of the various parameters is needed as a
consistency check of the SM. Moreover, SM processes can be directly sensitive
to new physics, they will allow to test QCD predictions and measure PDFs and
are backgrounds for many new physics channels. In this report, the status
of some analysis at ATLAS and CMS is reported focusing on W and Z inclusive
cross sections and on W and top quark mass measurements. Expected results on early
data are also shown. 
}

\section{Introduction}\label{subsec:prod}
The Large Hadron Collider (LHC) is a proton proton collider designed
to provide collisions at a center of mass energy of 14~TeV with a
nominal luminosity of 10$^{34}$~cm$^{-2}$s$^{-1}$.
The LHC was built at CERN as a discovery machine with the target to find the Higgs Boson and 
any possible evidence of new physics beyond the Standard Model (SM).
Two general purpose experiments (ATLAS~\citelow{ATLASDetPaper} and CMS~\cite{CMSTDR1}) have been constructed to study the collisions
provided by the LHC. They are currently in the final commissioning phase \cite{ATLAScomm,CMScomm} and are ready to collect data.

Before any discovery can be claimed a set of key issues should be addressed by the experiments:
the detectors response should be understood in detail and SM processes (W,Z,t) should be accurately 
measured as they can be considered as benchmark processes to assess the comprehension of the measurements.

The LHC can provide stringent tests of the Standard Model consistency measuring several fundamental parameters.
A non exhaustive list includes the W boson mass and width (M$_{W}$ and $\Gamma_{W}$ through the W boson 
decay distributions), the top quark mass 
m$_{t}$ and sin$^2$$\theta_{W}$ via the Z forward-backward asymmetry. 
In addition to that, some processes (e.g. rare top decays) have a direct sensitivity to new physics. 
Cross sections which are known theoretically with high accuracy (as those of the vector bosons) are
crucial to test QCD predictions in an unexplored regime and to measure Parton Density Functions (PDFs).
Moreover, SM processes should be carefully studied at the LHC because they are a background to 
many new physics channels.



At the time of writing these proceedings, April 2009, it is foreseen to have the first proton-proton 
collisions before the end of the year. It is planned to run without long intervals during the winter
with the goal to integrate $\sim$200~pb$^{-1}$. The energy per beam will be of 450~GeV at the beginning 
and will rise up to 5~TeV.

In this paper, the expected results for some measurements of SM processes at the LHC
are presented to show the status of the analyses and the expected performance.
More details and the description of the analyses not shown here can be found in \cite{ATLASCSC,CMSTDR2} 
and in the references cited in the text.
The studies are based on detailed simulations of the physics processes and of the detectors. 
The inclusive Z and W cross sections measurements are discussed in section \ref{wzxsec}; 
the measurement of the W mass is described in section \ref{wmass} and 
the top quark pair production cross section and top mass measurements are presented in section \ref{top}.

\section{Inclusive W and Z cross sections measurements.}\label{wzxsec}
The study of the production of W and Z events at the LHC is fundamental in several respects. First,
the calculation of higher order corrections is very advanced,
with a small theoretical uncertainty ($<$1\%). 
Such precision makes W and Z production a stringent test of QCD.

Z and W production cross sections are expected to be very large. For a center of mass energy
of 10~TeV~(14~TeV) we will have $\sigma$$_{W\rightarrow l\nu}$=14.3~nb~(20.5~nb) and
$\sigma$$_{Z\rightarrow ll}$=1.35~nb~(2.02~nb); the calculations are at NLO accuracy.
The experimental signatures for these
processes are very clean (in particular Z$\rightarrow ll$).
Thus, they will be extensively used as $standard$ $candles$ processes for understanding the experiments 
and tuning the Montecarlo, providing calibration and alignment of the detectors; setting the energy scales 
and resolutions and measuring the efficiencies for leptons.

Finally, the clean and fully reconstructed leptonic final states, in Z events, will allow a precise measurement 
of the transverse momentum and rapidity distributions. These distributions will constrain non
perturbative QCD aspects and the PDFs.
The high expected statistics will bring significant improvement on all these aspects, and this
improvement translates to virtually all physics at the LHC, where strong interaction and PDF uncertainties 
are a common factor.

The selection of Z$\rightarrow\mu\mu$ events (the CMS analysis~\cite{CMSZWmu} is presented) starts requiring one single muon
at the trigger level. Two high p$_T$ muons (p$_T$$>$20.0~GeV) 
with opposite charge sign should be reconstructed and should be isolated ($\sum$p$_T$$<$3 GeV for all the other tracks 
in a cone $\Delta$R$<$0.3).

In the case of Z$\rightarrow$$ee$ (ATLAS analysis~\cite{ATLASCSC} is discussed), the trigger requires one electron with p$_T$$>$10~GeV. 
Two clusters in the Electromagnetic Calorimeter (E$_T$$>$15~GeV) are then required at reconstruction stage;
they should be isolated ($\sum$E$_T$/E$_T$$^e$$<$0.2 in a cone $\Delta$R$<$0.45, where the electron energy is excluded in $\sum$E$_T$). 
These analyses are based on robust cuts in order to be safe against harsh experimental conditions; for example a common vertex is not
required nor a cut based on the tracks impact parameter.

\begin{figure}[h]
\begin{center}
\begin{minipage}{2.9in}
\begin{center}
\psfig{figure=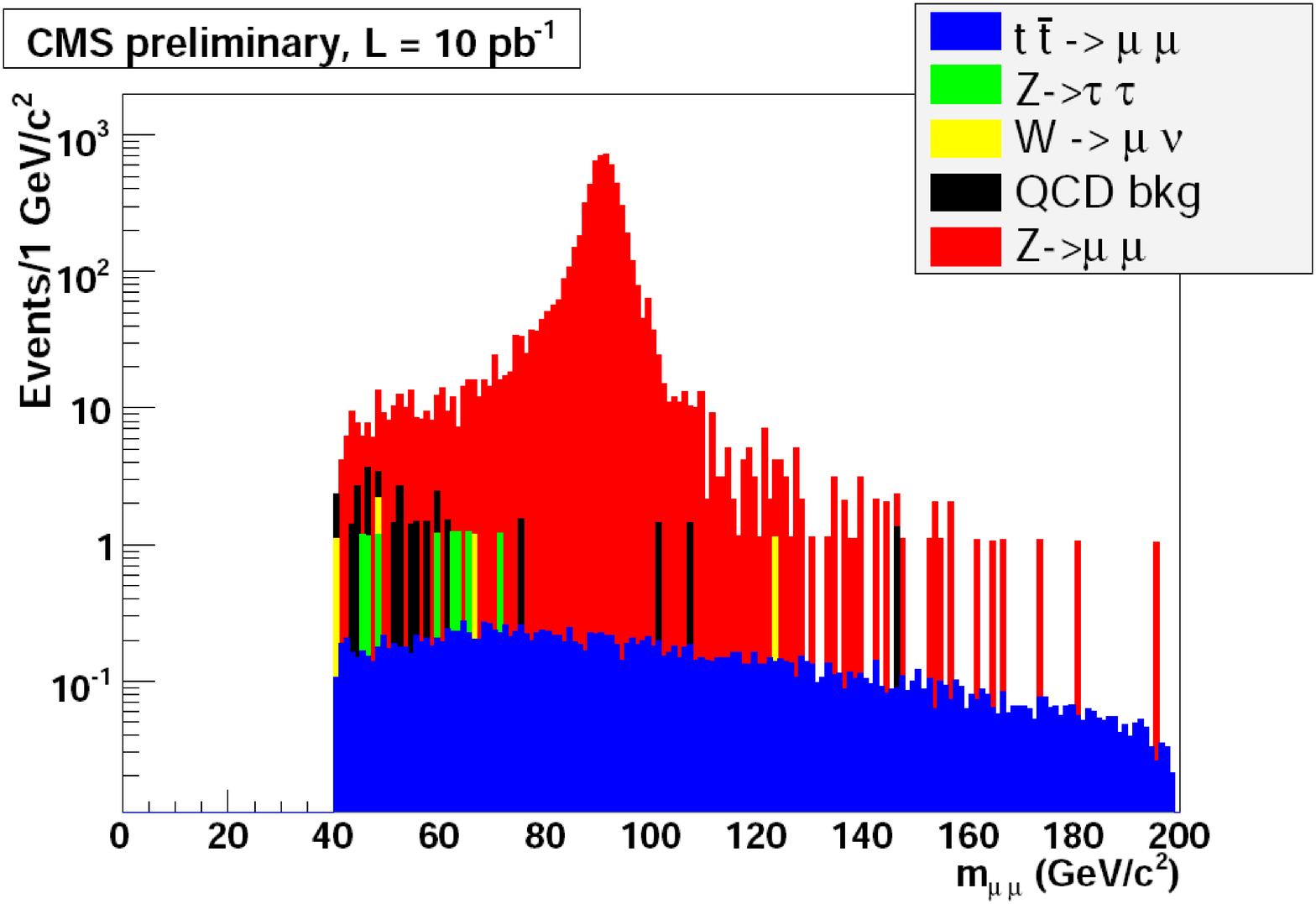,height=2.in}
\caption{Invariant mass distributions for Z$\rightarrow\mu\mu$ as expected in CMS 
for an integrated luminosity of 10~pb$^{-1}$
.
\label{zmumu}}
\end{center}
\end{minipage}
\hspace{0.1in}
\begin{minipage}{3.2in}
\begin{center}
\psfig{figure=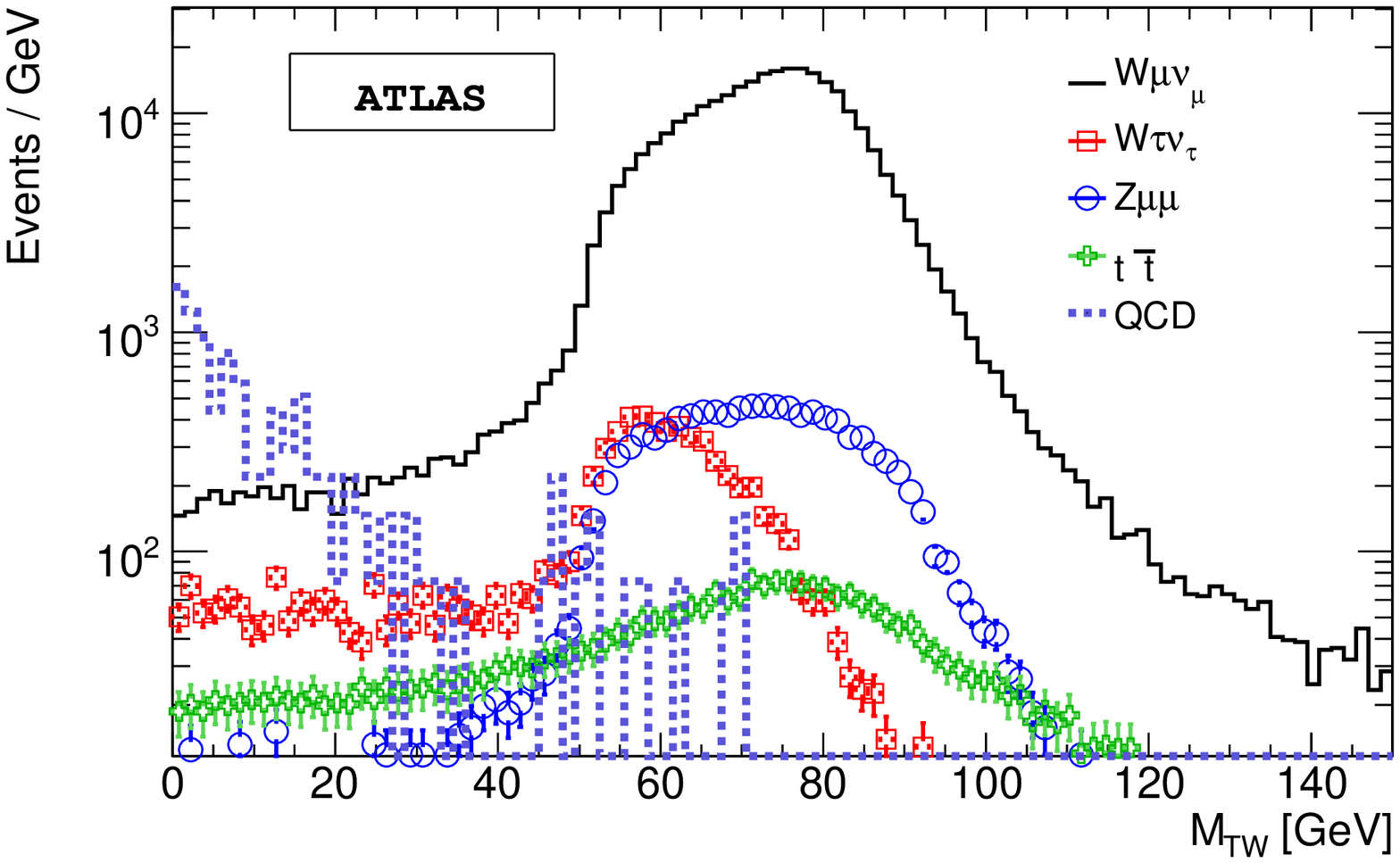,height=2.in}
\caption{Transverse mass distribution for W$\rightarrow\mu\nu$ events as expected in ATLAS for an integrated 
luminosity of 50~pb$^{-1}$ before the cut on M$_{TW}$ described in the text.
\label{wmunu}}
\end{center}
\end{minipage}
\end{center}
\end{figure}

The background estimation can be performed from the sidebands and/or from simultaneous fit to signal and background. 
Anyway this is a low background sample, in particular in the muon case.
The results of the selections are shown in figure~\ref{zmumu} for the case Z$\rightarrow\mu\mu$ channel.

The selection of W$\rightarrow$$e\nu$ events (CMS cuts~\cite{CMSZWe} are reported) starts requiring a single isolated electron at trigger level. 
A high E$_T$ electron (E$_{T}$$>$30~GeV) should be reconstructed; it is required to be isolated in the tracker
(no tracks with p$_{T}>$1.5~GeV, except for the one of the electron, within a cone $\Delta$R$<$0.6), in the electromagnetic calorimeter ($\sum$E$_T$/E$_T^{e}<$0.02; $\Delta$R$<$0.3)
and in the hadronic calorimeter ($\sum$E$_T$/E$_T^{e}<$0.10; 0.15$<$$\Delta$R$<$0.3). Events with a second electron
having E$_{T}>$20~GeV are rejected. 

W$\rightarrow\mu\nu$ events are obtained (in the ATLAS analysis~\cite{ATLASCSC}) with one muon having p$_T$$>$20~GeV at the trigger level.
A high p$_T$ muon (p$_T$$>$25~GeV) should be reconstructed and must be isolated (the energy deposited in the calorimeter around the muon track in a cone $\Delta$R$<$0.4 should be lower than 5~GeV).
Cuts on E$_T^{Miss}$$>$25~GeV and M$_{TW}$$>$40~GeV are also applied.

In the electron final state the dominant background are jet final state events.
To determine the shape of the background two method are foreseen; CMS propose to use 
a data sample passing electron selection with isolation criteria inverted while
ATLAS exploits a $\gamma$+jet sample obtained using the same selection as for the signal but requiring no matching tracks
in the Inner Detector.

In the muon final state the background comes from Z$\rightarrow\mu\mu$ and W$\rightarrow\tau\nu$ events. 
As these processes are well understood, the shape of the background will be obtained from Montecarlo.	
In figure \ref{wmunu} the transverse mass distribution is shown for the case W$\rightarrow \mu\nu$.


\begin{equation}\label{xseceq}
\sigma_{W(Z)}=\frac{N^{obs}_{W(Z)} - B_{W(Z)}}{\epsilon_{W(Z)}\cdot A_{W(Z)}\cdot\int\textit{L}dt} \hspace{0.5cm} ;\hspace{0.5cm}  
\frac{d\sigma_{W(Z)}}{\sigma}=\frac{dN\oplus dB}{N-B}\oplus\frac{d\epsilon}{\epsilon}\oplus\frac{dA}{A}\oplus\frac{dL}{L}
\end{equation}
The cross sections and the relative errors are computed as in eq.~\ref{xseceq}. $N^{obs}_{W(Z)}$ is the number of selected events; 
$B_{W(Z)}$ is the estimated number of background events, $\epsilon_{W(Z)}$ and $A_{W(Z)}$ are
respectively the trigger and reconstruction efficiency and the acceptance and $\int\textit{L}dt$ is the integrated luminosity.
The efficiencies will be computed from data using the Tag and Probe technique while the acceptance will be computed using the Montecarlo.
At the beginning the dominant uncertainty ($\sim$10$\%$) will come from luminosity measurements. This number will be reduced with a
better knowledge of the machine and when ALFA~\cite{ALFA} will come into play. 

The expected uncertainties on the measured cross sections are presented for different values of the integrated luminosity 
in table \ref{xsecrestab}. The error on the luminosity is not taken into account.
The uncertainty on identification and reconstruction efficiencies are expected to be at the level of 1\% (3\%) in the electron (muon) channel in the 
initial phase and to go well below 1\% on a longer time scale after an integrated luminosity of ~1~fb$^{-1}$ will be collected. 
The uncertainty on the background is foreseen to be at the level of 5\% ($<$1\%) in the electron (muon) channels at the beginning;
they will be reduced with more stringent selections when the available statistics increases.

\begin{table}[b]
\caption{Expected uncertainties on the measured cross sections; 
the error on the luminosity is not considered.\label{xsecrestab}}
\vspace{0.4cm}
\begin{center}
\begin{tabular}{|c|c|c|c|}
\hline
Experiment & Process & $\int\textit{L}dt$ & $\Delta\sigma$/$\sigma$~($\%$) \\
\hline
 
CMS &  pp$\rightarrow$Z+X$\rightarrow$$ee+X$ & 10~pb$^{-1}$ & 1.9~(stat)~$\pm$~2.3~(syst) \\
    &  pp$\rightarrow$W+X$\rightarrow$$e+X$  &            & 1.2~(stat)~$\pm$~5~(syst) \\

ATLAS &  pp$\rightarrow$Z+X$\rightarrow\mu\mu+X$ & 50~pb$^{-1}$ & 0.8~(stat)~$\pm$~3.8~(syst) \\
      &  pp$\rightarrow$W+X$\rightarrow\mu+X$    &              & 0.2~(stat)~$\pm$~3.1~(syst) \\

CMS & pp$\rightarrow$Z+X$\rightarrow\mu\mu+X$& 1~fb$^{-1}$ & 0.13~(stat)~$\pm$~2.3~(syst) \\
    & pp$\rightarrow$W+X$\rightarrow\mu+X$   &             & 0.04~(stat)~$\pm$~3.3~(syst) \\
ATLAS & pp$\rightarrow$Z+X$\rightarrow$$ee+X$  &             & 0.20~(stat)~$\pm$~2.4~(syst) \\
      & pp$\rightarrow$W+X$\rightarrow$$e+X$   &             & 0.04~(stat)~$\pm$~2.5~(syst) \\

\hline
\end{tabular}
\end{center}
\end{table}

\section{W mass measurement.}\label{wmass}
The W mass (M$_{W}$) is a fundamental parameter of the SM. It is related to the mass of the top quark and the mass of the Higgs 
boson and needs to be measured with highest precision. The LHC aims at improving the current world average (M$_W$=80399$\pm$25~MeV);
the W production cross section is 10 times larger than at Tevatron and the luminosity is higher.
W candidate events are selected as described in the previous section. The W mass can be extracted from the distribution of one 
of the two observables that are most sensitive to M$_{W}$: the transverse momentum of the lepton (p$_{T}^{l}$) and 
the transverse mass of the lepton-neutrino system (M$_{TW}$).
The value of M$_{W}$ is obtained fitting the measured distributions with template distributions.
The two analyses are complementary and are affected by different systematic effects. The shape of the p$_{T}^{l}$ distribution is
distorted by the transverse momentum of the W while M$_{TW}$ is mainly affected by the finite resolution of the detectors.
Z events are crucial in this analysis to build the templates and to properly account for experimental quantities like
the lepton energy scale, the energy resolution and the reconstruction efficiency.
There are several approaches to generate the templates. In the ATLAS analysis, the distributions are generated with the Montecarlo
and then are convoluted with the momentum scales, resolutions and missing E$_T$ response measured in Z events.
The CMS collaborations uses two methods. The scaled observable method uses templates obtained by transformation of the distributions
of p$_{T}^{l}$ or of the Z transverse mass into the corresponding quantities for the W. 
The other method is a kinematic transformation on a event by event basis (Morphing Method):
the lepton momentum in Z rest-frame is rescaled taking into account the ratio M$_{W}$/M$_Z$, one lepton is removed to simulate the neutrino and the
observable is boosted back to detector frame.

The ATLAS collaboration studied the result of the analysis also in the case of a limited statistics; this is intended as a study to set the 
method and to understand what can be done with very early data. The results are listed in table~\ref{wmasstab}
for the different channels and observables. 

\begin{table}[ht]
\caption{Expected uncertainties on the W mass for 15~pb$^{-1}$ of data.\label{wmasstab}}
\vspace{0.4cm}
\begin{center}
\begin{tabular}{|c|c|c|c|c|}
\hline
             & p$_{T}^{e}$ & p$_{T}^{\mu}$ & M$_{TW}(e)$ & M$_{TW}(\mu)$ \\
\hline
 
Statistical~(MeV)  & 120         & 106           & 61         & 57 \\
Experimental~(MeV) & 114         & 114           & 230        & 230 \\
Theo (PDF)~(MeV)   & 25          & 25            & 25         & 25  \\  
TOTAL~(MeV)        & 167         & 158           & 239        & 238 \\    

\hline
\end{tabular}
\end{center}
\end{table}

When the statistics will increase the measurement will become competitive. 
The CMS analysis performed with 1~fb$^{-1}$ of data, for example in with the Morphing Method applied to the muon channel, results in a value of 
$\Delta$M$_{W}$=40~(stat)~$\pm$~64~(syst.exp)~$\pm$~20~(syst.theo)~MeV. The experimental error is dominated by the missing transverse energy scale
and resolution while the theoretical uncertainty is mostly related to the uncertainties in the PDFs. It has been estimated that the systematic 
uncertainties can be further improved with increasing statistics; the expected result for 10~fb$^{-1}$ is 
$\Delta$M$_{W}$=15~(stat)~$\pm$~30~(syst.exp)~$\pm$~10~(syst.theo)~MeV.

\section{Top Quark observation and mass measurement.} \label{top}
The top quark has been discovered at Tevatron and their current value for the top quark mass 
is M$_{t}$=173.1$\pm$0.6~(stat.)$\pm$1.1~(syst.)~GeV \cite{TEVTMASS}.
A more precise measurement of M$_{t}$ is needed for consistency checks of the Standard model and
to constrain the Higgs mass.
The LHC offers a great opportunity to this extent because top quark pair production cross section, mainly via gluon-gluon fusion,
is of 833~pb (at NLO for $\sqrt{s}$=14~TeV), two order of magnitudes larger then at Tevatron. This measurement will be soon limited by systematic effects.

The golden channel is the lepton+jets channel in which the W from t ($\overline{t}$) quark decays leptonically and the W from 
the $\overline{t}$ (t) quark decays in 2 jets: $t\overline{t}\rightarrow$$Wb$$+$$W\overline{b}\rightarrow$$(l\nu)b$$+$$(jj)\overline{b}$.
This channel can be selected with good purity (the isolated lepton is exploited for triggering); the hadronic side is used to measure 
the top mass. The dominant background are W/Z+jets events. Other background are $t\overline{t}$ in other channels and single
top events. Multi-jet background (with fake leptons and missing E$_T$) has a very large cross section and a tiny efficiency 
to the selection cuts; the simulation is difficult and data driven methods are needed to estimate this contribution. In any case, this is
expected to be lower than W+jets.

\begin{figure}[h]
\begin{center}
\begin{minipage}{3.3in}
\psfig{figure=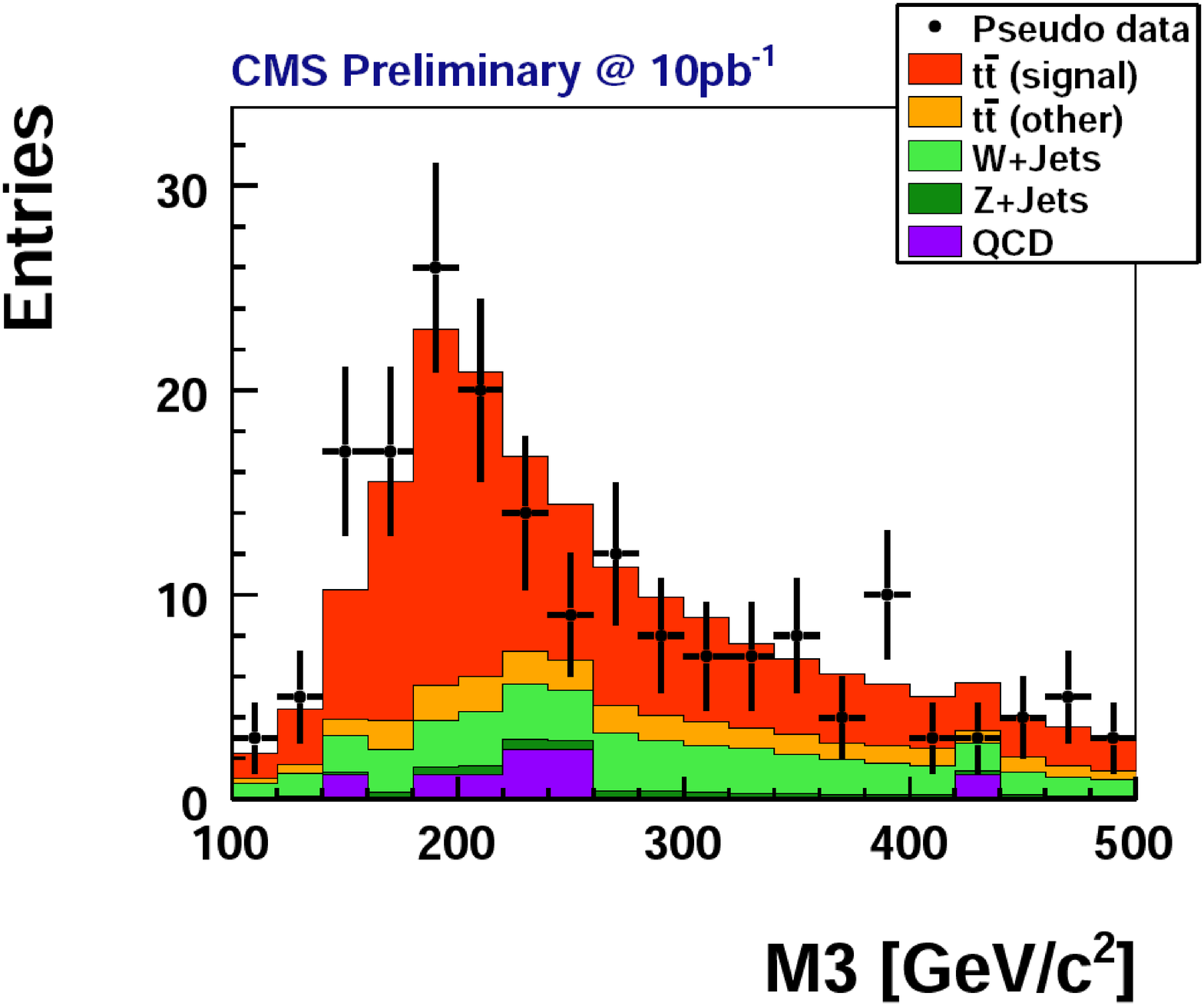,height=2.2in}
\caption{Invariant mass distributions of the three jets with the highest summed transverse energy as expected in CMS.
\label{topfigcms}}
\end{minipage}
\hspace{0.1in}
\begin{minipage}{2.8in}
\psfig{figure=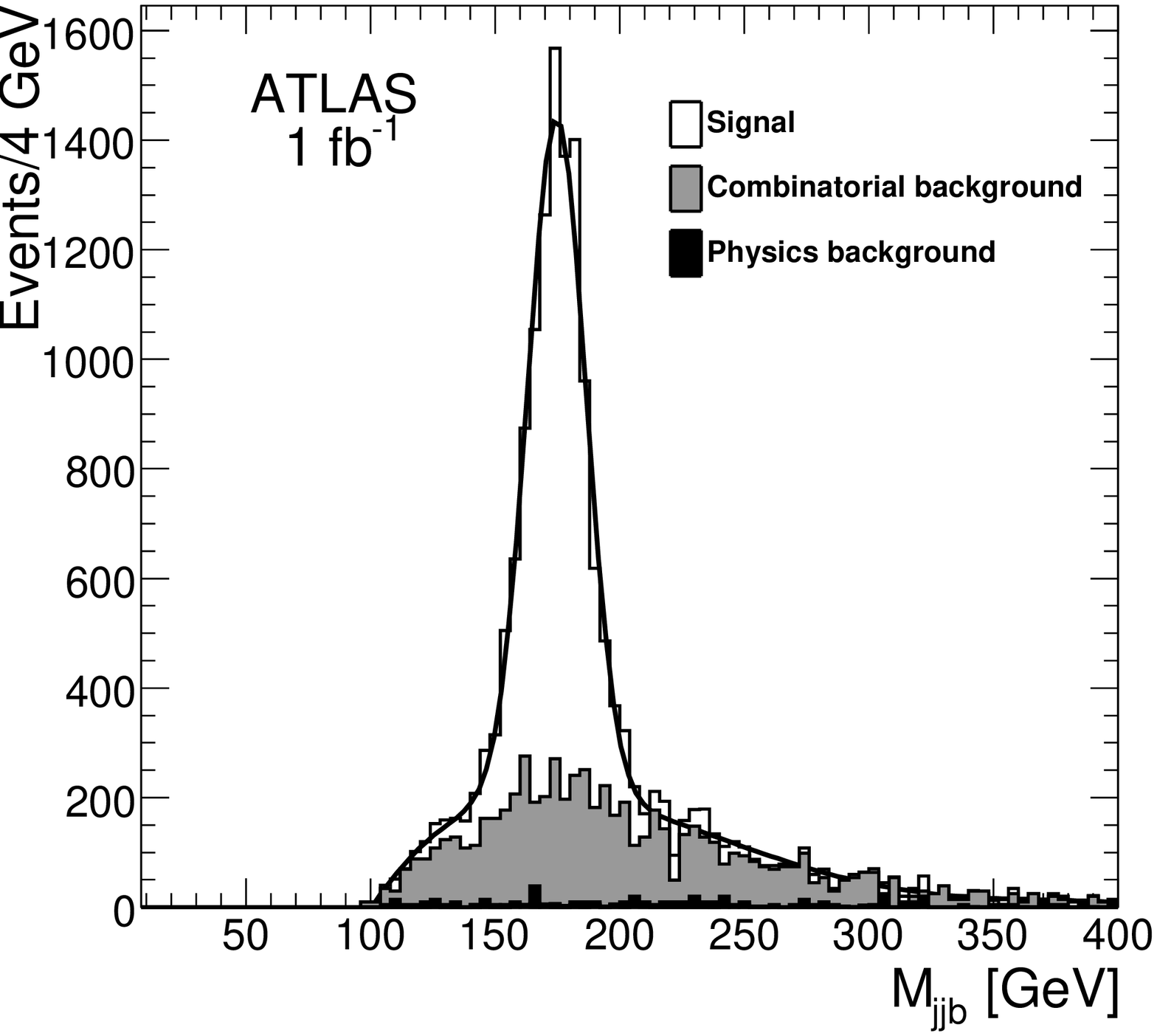,height=2.2in}
\caption{Invariant mass distributions of the three jets associated to the top quark as expected in ATLAS.
\label{topfigatlas}}
\end{minipage}
\end{center}
\end{figure}

The study of $t\overline{t}$ events require the reconstruction of many relevant experimental signatures
(e, $\mu$, jet, missing E$_T$, b-jet). Therefore the observation of the top signal will be a milestone in physics 
commissioning of the detectors. As an example, the expectations obtained by CMS \cite{CMStopobs} on 
the observation of the top signal with 10~pb$^{-1}$ is presented. The muon plus jets channel is selected
requiring a muon with p$_{T}$$>$30~GeV isolated in the calorimeters (the energy deposit in the calorimeter excluding the muon should be lower than 1~GeV) and in
the tracker (dR$_{min}$$>$0.3). At least 4 jets are required, three with E$_{T}$$>$40~GeV and
one with E$_{T}$$>$65~GeV. The cut on the number of jets highly reduces the QCD and W/Z+jets background.
No b-tagging of the jet is used. For 10~pb$^{-1}$ this results in a signal to background ratio
of 128/90. The overall selection efficiency (including the acceptance) is 10.3\%.
The shape of the W/Z+jets background is obtained from simulation and the normalization will come by comparison
with a control sample at low jet multiplicities.
The resulting invariant mass distribution of the three jets with the highest summed transverse energy is shown in figure~\ref{topfigcms}.

For more and more data being recorded and for improved understanding of the reconstruction performance, more refined analyses have been prepared; an example from ATLAS~\cite{ATLASCSC} is described in the following. The selection is based on
one isolated lepton with p$_T$$>$20(25)~GeV in the case of muons (electrons). A cut on E$_{T}^{Miss}$$>$20~GeV reduces the multi-jet
background while a cut on at least 4 jets with p$_{T}$$>$40~GeV rejects W/Z+jets events. Two of the jets are required to be tagged 
as b-jets.

The Jet Energy Scale (JES) is the main source of systematic uncertainties. Its effect is reduced by rescaling with a minimization
procedure. All possible light jet combinations are tried with a correction to their energy scale. The mass of a couple
is constrained to the W mass; the pair with the best $\chi^2$ is taken. To measure the top mass, the b-jet closest to the 
chosen pair is used. The invariant mass distribution of the three jets is shown in figure~\ref{topfigatlas} for an integrated
luminosity of 1~fb$^{-1}$. The effect of the light jet energy scale (reduced with the rescaling) is of 0.2~GeV/\%. The JES is expected 
to be known with a precision of 1\% with  1~fb$^{-1}$ of data. The uncertainty in the b-jet energy scale produces a larger uncertainty 
of 0.7~GeV/\%; it will be initially derived from the light JES using the Montecarlo and then complemented with Z+(b-jets) data.
The expected uncertainty on M$_{t}$ with 1~fb$^{-1}$ is of $\Delta$M$_{t}$=0.3~GeV~(stat.)$\pm$1~GeV~(syst.).

\section{Conclusions}
The LHC will start providing collisions late October this year. The first steps will be understanding the detector response and establishing
SM signatures. A strategy for the measurement of W and Z cross sections has been developed also for early data; simple and robust selections 
for electrons and muons have been set up to cope with the imperfections in calibration and alignment of the detectors.
The measurement of the W and top quark mass require a detailed detector understanding and will come at a later stage.
The Tag and Probe technique (applied on Z events) will provide the selection, reconstruction and trigger efficiencies directly from the data.
Some methods to estimate QCD backgrounds from data were developed. 
On a longer scale, for the precise measurement of SM parameters, the understanding of systematics is crucial.

\section*{Acknowledgments}
I would like to thank all the ATLAS and CMS colleagues who worked on the analyses and in particular the SM and Top Working group conveners for the discussions and for their advice in preparing this contribution.
I thank the organizers for the very interesting and pleasant conference.

%








\begin{thebibliography}{99}
\bibitem{CMSWmass} The CMS Collaboration, Prospects for the precision measurement of the W mass with the CMS detector at
the LHC, CMS NOTE 2006/061.
\end{thebibliography}
\end{document}